\documentclass[preprint]{elsarticle}
\usepackage{amsmath,amsfonts}
\usepackage{algorithmic}
\usepackage{array}
\usepackage{textcomp}
\usepackage{stfloats}
\usepackage{url}
\usepackage{verbatim}
\usepackage{graphicx}
\usepackage{hyperref}
\usepackage{multirow}
\usepackage{subcaption}

\def\BibTeX{{\rm B\kern-.05em{\sc i\kern-.025em b}\kern-.08em
    T\kern-.1667em\lower.7ex\hbox{E}\kern-.125emX}}
\usepackage{balance}
\begin{document}
\begin{frontmatter}

\title{A Semi-Supervised Algorithm for Improving the Consistency of Crowdsourced Datasets: The COVID-19 Case Study on Respiratory Disorder Classification}

\author[inst1]{Lara Orlandic\corref{fn1}}
\author[inst1,inst2]{Tomas Teijeiro}
\author[inst1]{David Atienza}

\cortext[fn1]{Corresponding author. \textit{E-mail address}: lara.orlandic@epfl.ch}

\address[inst1]{Embedded Systems Laboratory (ESL), EPFL, Lausanne, Switzerland}
\address[inst2]{Department of Mathematics, University of the Basque Country (UPV/EHU), Spain}

\begin{abstract}
Cough audio signal classification is a potentially useful tool in screening for respiratory disorders, such as COVID-19. Since it is dangerous to collect data from patients with such contagious diseases, many research teams have turned to crowdsourcing to quickly gather cough sound data, as it was done to generate the COUGHVID dataset. Furthermore, the COUGHVID dataset enlisted expert physicians to diagnose the underlying diseases present in a limited number of uploaded recordings. However, this approach suffers from potential mislabeling of the coughs, as well as notable disagreement between experts. In this work, we use a semi-supervised learning (SSL) approach -- based on state-of-the-art audio signal processing tools and interpretable machine learning models -- to improve the labeling consistency of the COUGHVID dataset and the robustness of COVID-19 versus healthy cough sound classification. First, we leverage existing SSL expert knowledge aggregation techniques to overcome the labeling inconsistencies and label sparsity in the dataset. Next, our SSL approach is used to identify a subsample of re-labeled COUGHVID audio samples that can be used to train or augment future cough classification models. The consistency of the re-labeled data is demonstrated in that it exhibits a high degree of class separability -- 3x higher than that of the user-labeled data -- despite the expert label inconsistency present in the original dataset. Furthermore, the spectral differences in the user-labeled audio segments are amplified in the re-labeled data, resulting in significantly different power spectral densities between re-labeled healthy and COVID-19 coughs in the 1-1.5 kHz range ($p=1.2 \times 10^{-64}$), which demonstrates both the increased consistency of the new dataset and its explainability from an acoustic perspective. Finally, we demonstrate how the re-labeled dataset can be used to train a cough audio signal classifier. Although we focus on the novel field of COVID-19 cough sound classification, the SSL approach can be used to combine the medical knowledge of several experts to improve the database consistency for any diagnostic classification task.
\end{abstract}


\end{frontmatter}

\section{Introduction}

Audio signal processing has the potential to become a useful diagnostic tool, particularly in the field of respiratory disorder detection \cite{rao_acoustic_2019}. Studies have shown that respiratory conditions such as pertussis, asthma and pneumonia can be automatically diagnosed using algorithms that analyze the cough sounds produced by patients \cite{Pramono2016, Amrulloh2015}. The benefits of using audio processing and Machine Learning (ML) algorithms to classify cough sounds is that the diagnosis can be performed quickly and easily by a device such as a smartphone, thus reducing the workload of medical professionals and supporting ubiquitous analysis. Since one of the most common symptoms of the novel coronavirus disease (COVID-19) is a dry cough \cite{whochina}, there has been significant interest in leveraging such algorithms to quickly and unobtrusively screen for the virus \cite{xia_covid-19_nodate, nessiem_detecting_2021, laguarta_covid-19_2020}.

In order to train ML algorithms to screen for COVID-19 from cough sounds, a large amount of cough sound data is necessary. However, since the disease is a highly contagious airborne pathogen \cite{lotfi_covid-19_2020}, collecting such cough sound data from COVID-19 positive individuals requires significant effort and sanitary precautions to ensure the safety of those involved. To overcome such data collection limitations, several groups around the world have relied on crowdsourcing, a paradigm in which internet users upload their own cough sounds and report whether or not they had been diagnosed with the condition \cite{xia_covid-19_nodate, laguarta_covid-19_2020, orlandic_coughvid_2021, sharma_coswara_2020}. However, such user-labeled data may suffer from mislabeling, as some extensive datasets reported that thousands of uploaded recordings did not contain cough sounds at all \cite{xia_covid-19_nodate, orlandic_coughvid_2021}. 

As a validation step of the crowdsourced recordings, the COUGHVID dataset enlisted four expert physicians to listen to a number of cough recordings and diagnose any audible respiratory disorders (ex. COVID-19, upper and lower respiratory infections) \cite{orlandic_coughvid_2021}. However, the general trend was that the four experts did not agree on the COVID-19 diagnosis. Disagreement between physicians is common in the medical field; a study of medical referrals noted that only 12\% of final diagnoses agreed with the initial diagnoses, and 21\% of final diagnoses significantly differed from the initial ones \cite{van_such_extent_2017}. Therefore, extra care must be taken to overcome the label ambiguity of any crowdsourced cough audio databases, as the expert disagreement and user mislabeling can lead to erroneous classification.

The winners of the 2017 PhysioNet/CinC Challenge on ECG signal classification observed that expert annotation inconsistencies in physiological data can be alleviated through manual re-labeling, thus leading to significant improvements in classifier performance on unseen data \cite{teijeiro_abductive_2018}. However, manual re-labeling of cough sounds is difficult to perform without extensive medical training. Furthermore, manually re-labeling such an extensive dataset would require significant time and effort. Semi-supervised learning (SSL) is a ML paradigm that can be used to automate the re-labeling process of biomedical signals \cite{zhu_speech_2021, deng_semisupervised_2018}. While SSL is often used in conjunction with Deep Learning models for medical inference \cite{zhu_speech_2021, guan_who_2018, raghu_direct_nodate, tanno_learning_2019}, it can also be used with classical ML approaches in which the extracted features leverage domain knowledge to shed light on the inner-workings of the classifier.

In this work, we utilize a state-of-the-art SSL technique based on explainable ML models that integrates knowledge of a variable number of human annotators. This technique uses the cough sound recordings of the COUGHVID dataset that were labeled by expert physicians to train three classifiers, where each one models the medical knowledge of a different expert. Next, we overcome the issue of expert label scarcity by generating pseudo-labels on the entire database using each expert model. Then, the outcomes of these models were compared alongside crowdsourced user labels to identify a subset of cough recordings with the highest probability of originating from either COVID-19 positive or healthy individuals. Thus, we overcome the issues of crowdsourced data mislabeling and expert label inconsistency by identifying a high-quality subsample of datapoints -- with a threefold increase in feature separability compared to the user-labeled data, as well as a more significant difference in the power spectral densities of the two cough classes ($p = 1.2 \times 10^{-64}$) -- which can be used to train future cough classifiers.

The subsample of cough audio recordings identified through our SSL approach was subsequently made available to the public for further classifier development and ML exploration. To assess the intra-class consistency of this data, we quantify the class separability of standard audio features extracted from COVID-19 versus healthy coughs in the SSL labeling scheme compared to that of the expert labels and crowdsourcing labels of the COUGHVID dataset. Finally, we demonstrate how this data can be used to train cough audio signal classifiers by training a final COVID-19 cough detection model and comparing its classification accuracy to that of the fully supervised models. As a result, in addition to applying SSL to the relatively novel task of COVID-19 screening from cough sounds, this work aims to provide an automated approach for increasing the labeling quality of biosignal datasets, which can be applied to many other pathologies.

\section{Related Works}

While most cough classification algorithms focus on fully supervised ML approaches \cite{Pramono2016, xia_covid-19_nodate, laguarta_covid-19_2020}, semi-supervised learning (SSL) has scarcely been applied to the task \cite{xue_exploring_2021}. In this paradigm, unlabeled data is exploited in augmenting the dataset to enhance the performance of the classifier, thus providing ample training data and overcoming the issue of label sparsity \cite{lee2013pseudo, zhang_semi-supervised_2012}. Semi-supervised audio classification algorithms have outperformed fully-supervised models both for audio signal categorization \cite{zhang_semi-supervised_2012} and cough detection \cite{hoa_semi-supervised_2011} tasks. Furthermore, Han et al. found that incorporating SSL into sound classification enabled a reduction of 52.2\% in human annotations necessary to achieve comparable results to fully-supervised methods \cite{han_semi-supervised_2016}.

A simple, state-of-the-art SSL technique is Pseudo-Label, in which an initial model trained on the labeled samples classifies the unlabeled samples, which are then assigned pseudo-labels based on which predicted class has the highest probability \cite{lee2013pseudo}. Then, a final model is trained using both the original labels and pseudo-labels as ground truth. While this method shows significant performance gains over classical supervised learning on the benchmark MNIST dataset, it does not address the issue of label ambiguity present when the labels of different annotators do not match.

In addition to overcoming label scarcity, SSL approaches have also proven successful in alleviating the burden of inconsistent, ambiguous, and erroneous labels on ML classification tasks \cite{zhou_brief_2018}. Considering the example of 3D image segmentation tasks, semi-supervised models have been shown to outperform fully supervised ones both in the presence of human mislabeling and added random noise \cite{lv_semi-supervised_2012}. Furthermore, SSL has been widely utilized in speech emotion recognition, a field that suffers from sparse, inconsistent labeling by multiple untrained annotators \cite{deng_semisupervised_2018, zhu_speech_2021}. In particular, Zhu et al. devised an iterative, semi-supervised scheme using the ambiguous emotion annotations of six to twelve annotators and concluded that sufficient training data and moderately reliable labels at the onset of training can significantly improve the classification performance with respect to fully supervised training \cite{zhu_speech_2021}.

Recent works leveraging SSL to overcome inconsistencies in expert physicians' labels utilize an approach in which each expert is modeled by a Deep Neural Network, and then the outputs of these expert models are combined to generate a final label for each sample \cite{guan_who_2018, raghu_direct_nodate, tanno_learning_2019}. For example, Li et al. applied this approach to electronic medical record entity recognition by training five distinct models, expanding them to the whole dataset using Pseudo-Label, and then using a majority voting algorithm to generate the final labels \cite{li_semi-supervised_2021}. Furthermore, Guan et al. used individual expert modeling for diabetic retinopathy classification and found that this approach outperformed the Expectation-Maximization algorithm traditionally used for weighing the accuracies of multiple raters \cite{dawid_maximum_1979}. This promising SSL expert modeling technique has not yet been applied to the field of cough audio signal classification.

Our approach leverages the insights from previous works on SSL to overcome the labeling inconsistencies of the COUGHVID dataset and identify a subset of cough audio samples with consistent labels. We use a pseudo-label-based SSL scheme to expand the expert labels onto unlabeled segments of the dataset, and then combine the expert labels to identify the final dataset. Similarly to \cite{zhu_speech_2021}, we analyze the trade-offs between label consistency and training data size when selecting the final SSL approach. As opposed to previous works that rely on Deep Learning \cite{zhu_speech_2021, deng_semisupervised_2018, guan_who_2018, raghu_direct_nodate, tanno_learning_2019, li_semi-supervised_2021}, which may be difficult to interpret and therefore not amenable to sensitive medical classification tasks, we rely on classical ML algorithms using state-of-the-art audio feature computation. Furthermore, we determine the importance of the various features to the classification outcome of the SSL approach versus the user or expert label based models to assess the similarities and differences between the approaches.

\section{Methods}

\subsection{Methodology Overview}

\begin{figure}[ht]
  \centering
  \includegraphics[width=0.4\linewidth]{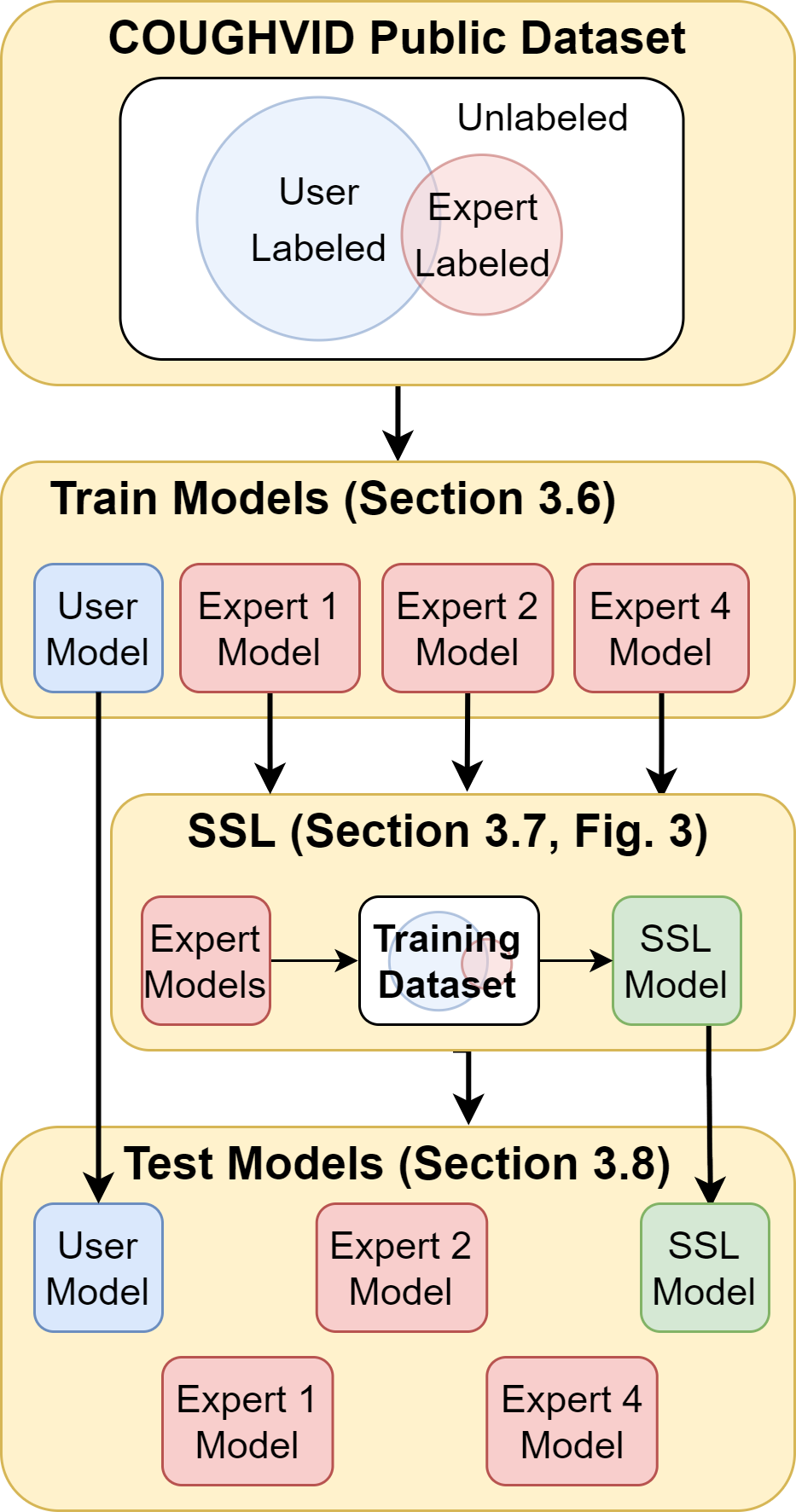} 
\caption{An illustration of the model development methodology, showing the different subsets of the COUGHVID dataset used at each stage. The supervised training, semi-supervised learning, and testing procedures are described in Sections \ref{sec:ml_opt}, \ref{sec:ssl}, and \ref{sec:testing}, respectively.}
\label{fig:methodology}
\end{figure}

One of the challenges of performing COVID-19 classification based on user-labeled and expert-annotated data is label ambiguity. Since the COUGHVID dataset is crowdsourced, it cannot be known with absolute certainty if the cough recordings labeled as COVID-19 or healthy truly originated from people with the condition or lack thereof. Furthermore, the experts' cough diagnoses exhibited a Fleiss' Kappa score of 0.07 \cite{Fleiss1971}, meaning that there was only a slight agreement between the four experts about the cough diagnoses \cite{orlandic_coughvid_2021}. 

As shown in Fig. \ref{fig:methodology}, we assessed the label consistency in each of the COVID-vs-healthy classification schemes provided by the dataset (i.e., users, experts) by extracting audio signal features and training ML models based on each set of labels. Then, the semi-supervised learning (SSL) approach was employed to produce a final classifier. Therefore, the following ML models were developed and compared in terms of various classification accuracy metrics on their respective labeling schemes:

\begin{enumerate}
    \item \textit{User Crowdsourcing Model:} In this classifier, the recordings in the positive class were self-labeled as ``COVID-19" by the users who uploaded them. Similarly, the negative class recordings were labeled as ``healthy" by the users.
    \item \textit{Expert [1,2,4] Model:} Three separate models were developed, corresponding to the labels of Experts 1, 2, and 4. Since we see from Table \ref{tab:label_counts} that Expert 3 only labeled one recording as COVID-19, this is not enough information for a ML model to reliably perform generalization. Therefore, this expert's labels are omitted from consideration in further analysis. The positive class was made up of recordings labeled by each expert as ``COVID-19", and the corresponding negative class was labeled as ``healthy\_cough".
    \item \textit{SSL Model:} In order to combine the knowledge from both the users and the experts into one model, semi-supervised learning was used. In this approach, the expert models and user labels were used to filter the dataset and determine the subset of coughs with the highest probability of being COVID-19 positive and healthy. The details of the implementation are described in Section \ref{sec:ssl}.
\end{enumerate}

\subsection{Dataset Description}
\label{sec:coughvid}

This analysis uses the COUGHVID crowdsourcing dataset, which is a vast repository of cough audio samples originating from diverse participants located across the globe \cite{orlandic_coughvid_2021}. The dataset is made up of user-uploaded cough recordings, many of which contain a status label indicating whether the user claimed to be diagnosed with COVID-19, exhibiting symptoms, or healthy at the time of recording. As an additional validation step, four expert physicians each labeled 1,000 cough recordings to diagnose potential respiratory disorders (i.e., COVID-19, upper respiratory infection) that are audible in the recordings. Each expert reported spending approximately 10 hours to label the cough sounds, which exemplifies the significant time and effort human labeling takes for such a task.

An expanded version of the training dataset was used, containing recordings uploaded from April 2020 to October 2021. There are about 34,500 recordings in this dataset, 20,644 of which contain user status labels. Both the expert labels and the testing dataset described in \cite{orlandic_coughvid_2021} are unchanged in this work. Table \ref{tab:label_counts} displays the value counts of the cough sounds labeled as COVID-19-positive and healthy by the users and each of the experts. It should be noted that the coughs labeled by the users and experts are not mutually exclusive, and 150 coughs were annotated by all experts to assess the level of agreement between the physicians.

\subsection{Cough Audio Signal Pre-Processing}

Since the COUGHVID dataset contains some recordings that do not capture cough audio, the cough classifier developed in \cite{orlandic_coughvid_2021} was used to remove non-cough recordings from consideration. Furthermore, only recordings with a cough classifier output greater than 0.8 were used in this work.

\begin{table}
\centering
\caption{Recording Counts in the COUGHVID Training Dataset}
\begin{tabular}{|l|l|l|} 
\hline
Label Origin & Healthy & COVID-19  \\ 
\hline
Users         & 15,476  & 1,315     \\ 
\hline
Expert 1     & 259     & 279       \\ 
\hline
Expert 2     & 67      & 285       \\ 
\hline
Expert 3     & 199     & 1         \\ 
\hline
Expert 4     & 221     & 84        \\
\hline
\end{tabular}
\label{tab:label_counts}
\end{table}

As an initial pre-processing step, all of the cough recordings were normalized to their maximum absolute value such that the signal values range from -1 to 1. This enables a fair comparison of the RMS power of different signal segments and provides numerical stability. Next, a 4th order Butterworth lowpass filter with a cutoff frequency of 6 kHz was applied. Consequently, the recordings were downsampled to 12 kHz. This filtering was performed to reduce high-frequency noise and increase the computational efficiency of all further signal processing and feature extraction algorithms. The cutoff frequency was chosen because visual analysis of the cough signal spectra revealed that most of the signal power lies below 6 kHz. Furthermore, past cough sound classification algorithms used cutoff frequencies ranging from 4 Hz to 8 Hz \cite{Pramono2016,Chatrzarrin2011,Drugman2011AssessmentPublication}, so an intermediate value was chosen for this work.

\subsection{Cough Segmentation}
\label{sec:cough_seg}

Once the recordings were pre-processed, a custom cough segmentation algorithm was employed to isolate each individual cough event present in a given recording. The segmentation algorithm exploits cough physiology to divide each recording into its constituent cough sounds. This algorithm enables feature extraction on each cough, thus suppressing silence and extraneous low-amplitude sounds like breathing. Furthermore, the algorithm can be used to perform a simple Signal-to-Noise Ratio (SNR) calculation, as well as aggregation of the ML classifier labels of all coughs originating from the same recording.

\begin{figure}[ht]
  \centering
  \includegraphics[width=\linewidth]{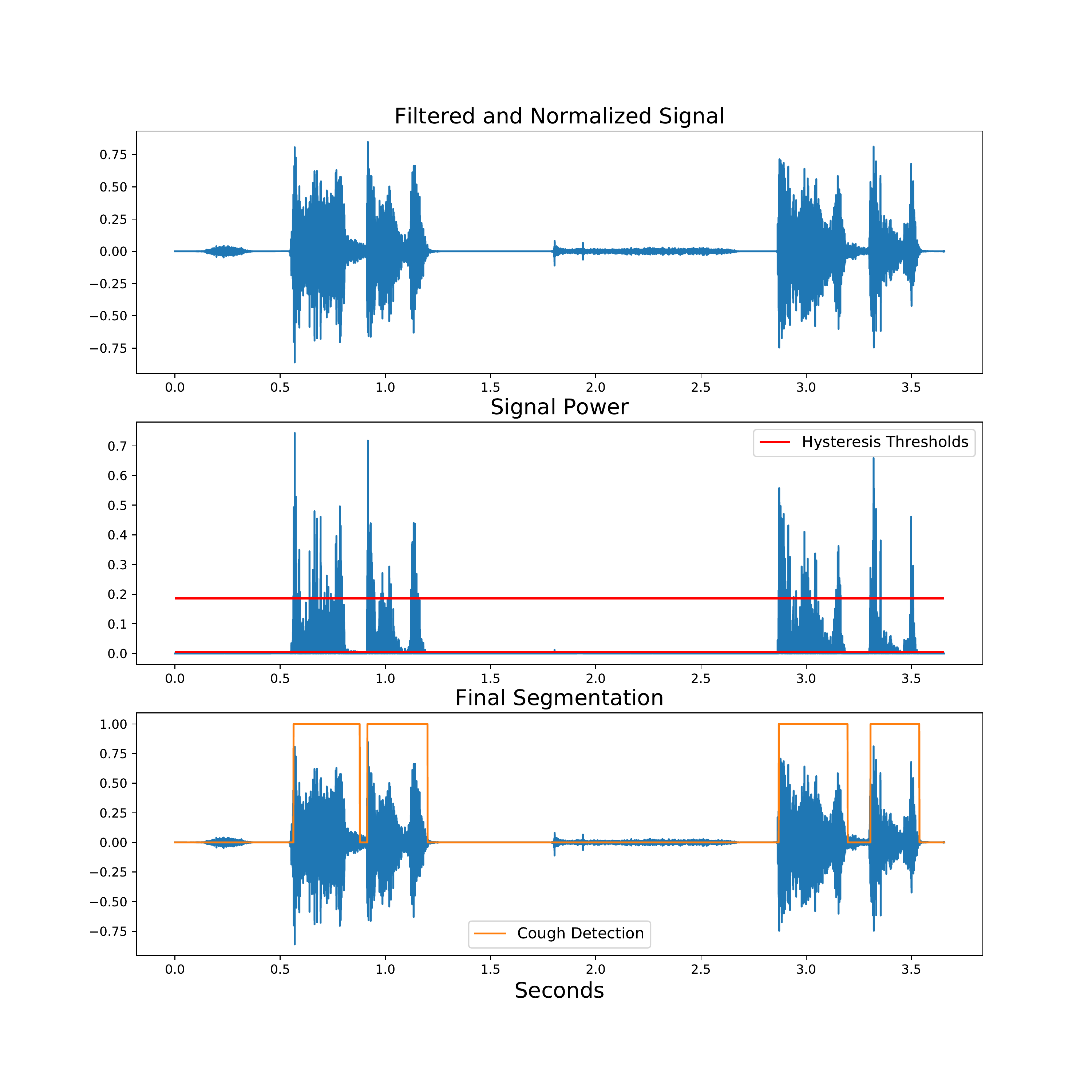} 
\caption{A step-by-step illustration of the cough segmentation procedure. A hysteresis comparator was applied to the signal power to detect the sound bursts, which enables us to consistently discard the segments without relevant inputs for the subsequent ML process.}
\label{fig:cough_seg}
\end{figure}

The algorithm is depicted in Fig. \ref{fig:cough_seg} on a recording of a breath, two coughs, another breath, and two more coughs. First, the signal is squared to compute its power. Next, a hysteresis comparator is applied to extract the sudden bursts in sound amplitude that arise from coughing. This means that potential cough candidates are determined to be regions started by the signal exceeding the upper threshold and ended by the signal going below the lower threshold. A tolerance of 10 ms is applied to the thresholds, meaning that the signal should either exceed the upper threshold or go below the lower threshold for at least 10 ms for a cough onset or offset to be recorded. The lower and upper hysteresis thresholds were set to 0.1 and 2 times the RMS signal power, respectively. These multipliers were empirically determined through analysis of a variety of cough audio signals. 

Next, the cough segments were analyzed and discarded based on the physiological limitations of cough duration. The cough is composed of three segments: inspiration, compression, and expiration. The latter two have known timing constraints: the compressive phase, during which inhaled air is compressed in the lungs to increase lung pressure, typically lasts 200 ms \cite{Chang2006TheCough}. The expiratory phase is initiated by a brief opening of the glottis (30-50 ms), causing the loudest phase of the cough sound and rapid airflow, followed by 200-500 ms of lower respiratory airflow \cite{Chang2006TheCough}. Therefore, the minimum possible cough sound length is approximately 230 ms. We consequently discard any cough sound candidates shorter than 200 ms, and we include the 200 ms before and after the cough candidate in each segmented cough to capture any low-amplitude noise caused during the compressive and expiratory phases. 

The cough segmentation algorithm was subsequently used to eliminate cough recordings with significant background noise. An estimate of the SNR was calculated for each signal as described in \cite{orlandic_coughvid_2021} by comparing the RMS signal power of the cough segments of a recording to that of the non-cough segments. The training and testing datasets were further filtered by retaining only the recordings with a SNR greater than 5, above which cough sounds were clearly more prominent than background noise. The cough segmentation and SNR estimation algorithms are available in the \href{https://c4science.ch/diffusion/10770/}{COUGHVID public git repository}\footnote[1]{https://c4science.ch/diffusion/10770/} to foster reproducibility.

\subsection{Feature Extraction}
\label{section:feature_extraction}
The set of 60 features extracted from each cough audio segment are the same as those computed in the cough classification algorithm in \cite{orlandic_coughvid_2021}. These features are a mixture of time, frequency, and Mel frequency domain computations that were chosen due to their previous implementation in automatic cough sound classification ML algorithms \cite{Pramono2016, Chatrzarrin2011}. These features provide both general information about the signal spectra, as well as detailed computations regarding specific frequency bands, thus allowing the subsequent feature elimination step to select only the relevant features to each classification task. The code used for feature extraction is available to the public in the COUGHVID repository \cite{orlandic_coughvid_2021}. 

For each classification task described in Section \ref{sec:ml_opt}, an additional set of Power Spectral Density (PSD) features were selected by inspecting the averaged PSD of each class in the training dataset, divided by the total average signal power such that the PSD curve is normalized to a unit area. The frequency bands displaying a large variation between the average normalized PSDs of the two classes were noted, and the bandpowers within these frequency ranges were added as features. This produced a variable number of PSD features for each classifier.

Some of the user-labeled data in the COUGHVID dataset contains user metadata information, such as their reported age, gender, and presence of respiratory disorders. In order to provide the model with some user-specific information to assist in  classification, the binary gender value was added to the feature set. In case no gender information was provided, a gender identification model was developed using the ML model optimization procedure described in Section \ref{sec:ml_opt}. The model was trained using the training data subset containing gender labels, and resulted in a classifier with an area under the receiver operating characteristic curve (AUC) of 0.8 on the testing dataset described in Section \ref{sec:testing}. This classifier was used to assign a gender of ``male" or ``female" to any cough in the dataset for which this value was not provided.

\subsection{Model Comparison and Optimization}
\label{sec:ml_opt}

For each classification task described in Section \ref{sec:coughvid}, a ML model was trained to distinguish COVID-19 from healthy coughs based on the feature vectors of the training dataset. 
Prior to optimization, the features were standardized by removing the mean and scaling to unit variance.

Next, we compared the efficacy of seven different state-of-the-art binary classification ML algorithms: Logistic Regression (LR), K Nearest Neighbors (KNN), Decision Tree Classifier (DTC), Gaussian Naive Bayes (GNB), Random Forests (RF), eXtreme Gradient Boosting (XGB), and Linear Discriminant Analysis (LDA), most of which were implemented in the Python scikit-learn library \cite{noauthor_scikit-learn_nodate}. To ensure a fair comparison between the different algorithms, the hyperparameters of each model were tuned simultaneously using Tree-structured Parzen Estimates (TPE) \cite{NIPS2011_4443}. The objective of the TPE procedure was to find the combination of hyperparameters that produced the highest mean area under the receiver operating characteristic curve (AUC) across 5 cross-validation (CV) folds.

The utilized CV procedure was a 5-fold GroupShuffleSplit \cite{noauthor_sklearnmodel_selectiongroupshufflesplit_nodate}; in each CV fold, 20\% of the recordings were randomly selected and used for validation, and the remaining recordings were used for training. The segmented coughs that comprised these recordings were correspondingly assigned to training or validation. This ensured that no coughs originating from the same recording were included in both the training and validation sets of each fold, thereby maintaining the generalizability of our results to unseen cough recordings.

As shown in Table \ref{tab:label_counts}, there is a significant class imbalance in each of the classification tasks. This issue was addressed using the Synthetic Minority Over-Sampling Technique (SMOTE) \cite{chawla_smote_2002}, which was employed to generate synthetic training samples from linear combinations of the minority class in the training dataset. This procedure produced a balanced sample of COVID-19 and healthy labeled training coughs by creating synthetic samples of the minority class of the training data within each CV fold.

Following TPE, the final mean and standard deviation AUC scores of all of the optimized models were analyzed. The model with the highest mean AUC was chosen, and its learning curve was analyzed to determine if the model was underfitting or overfitting, and whether or not the results converged to a consistent performance with the amount of data available. In the case of overfitting, Recursive Feature Elimination with Cross-Validation (RFECV) was performed on the optimized model to recursively remove the weakest features of the model. This technique has the potential to reduce the variance of the model through the elimination of weak features, but risks increasing the bias of the model by potentially eliminating important features \cite{Munson2009OnBagging}. Finally, the same TPE procedure was used to re-optimize the hyperparameters of the model with a reduced feature set.

An advantage of cough segmentation is that it enables aggregation of the classifier outputs of coughs originating from the same recording, which potentially enhances the accuracy of the classifier. Each recording was segmented into $N$ cough sounds, and each cough was processed separately by the trained classifier. This resulted in a series of classifier output probabilities $[p_1, p_2, ... , p_N]$, corresponding to the probability that each cough signal is COVID-19 positive. Since this diagnosis cannot change from one cough to the next, the probabilities can be combined to form one classifier output per recording, $p_{total}$. We employed two different aggregation techniques, all based on the logit score of each cough:

\subsubsection{Logit Mean}
\begin{equation}
    p_{total} = \frac{1}{N} \sum_{i=n}^N log(\frac{p_i}{1-p_i})
    \label{eq:logit_mean}
\end{equation}

\subsubsection{Logit Median}
\begin{equation}
    p_{total} = Median[ log(\frac{p_i}{1-p_i}) ]
    \label{eq:logit_median}
\end{equation}

Once the optimized model was selected, one final CV split was generated to form a training and validation dataset. The model was trained on this reduced training dataset, and the ROC curve was plotted using both of the logit aggregation methods. The aggregation method with the highest AUC value for the validation set was selected. Furthermore, the optimal classifier decision threshold was determined for computing further accuracy metrics by selecting the aggregated logit threshold with the highest geometric mean between the model's sensitivity and specificity. These final hyperparameters were noted for use in testing, and the model was re-trained using the full training dataset. The final model was tested on the private, unseen COUGHVID test set, as described in Section \ref{sec:testing}.

\subsection{Semi-Supervised Learning (SSL)}
\label{sec:ssl}

Instead of relying solely on the potentially noisy user labels or the often contradictory expert labels described in Section \ref{sec:coughvid}, an SSL approach was used to overcome the issues of label inconsistency and ambiguity by identifying a subset of consistent training data with a high probability of belonging to COVID-19-positive or healthy subjects. At a high level, this method aims to distill the knowledge of each expert onto samples that the expert did not annotate, similarly to what is done in the state-of-the-art Pseudo-Label method \cite{lee2013pseudo}. Then, the agreement between the experts' models is used to identify a set of recordings with high label confidence, similarly to previous work on SSL applied to medical datasets with inconsistent labels \cite{guan_who_2018}. This is similar to the cross-voting methodology utilized by Li et al. \cite{li_semi-supervised_2021}, except that instead of randomly partitioning the data to train each model, the data is divided based on the expert that annotated it, thereby modeling each expert's medical expertise using ML.

\begin{figure*}[ht]
	\centering
	\includegraphics[width=\textwidth]{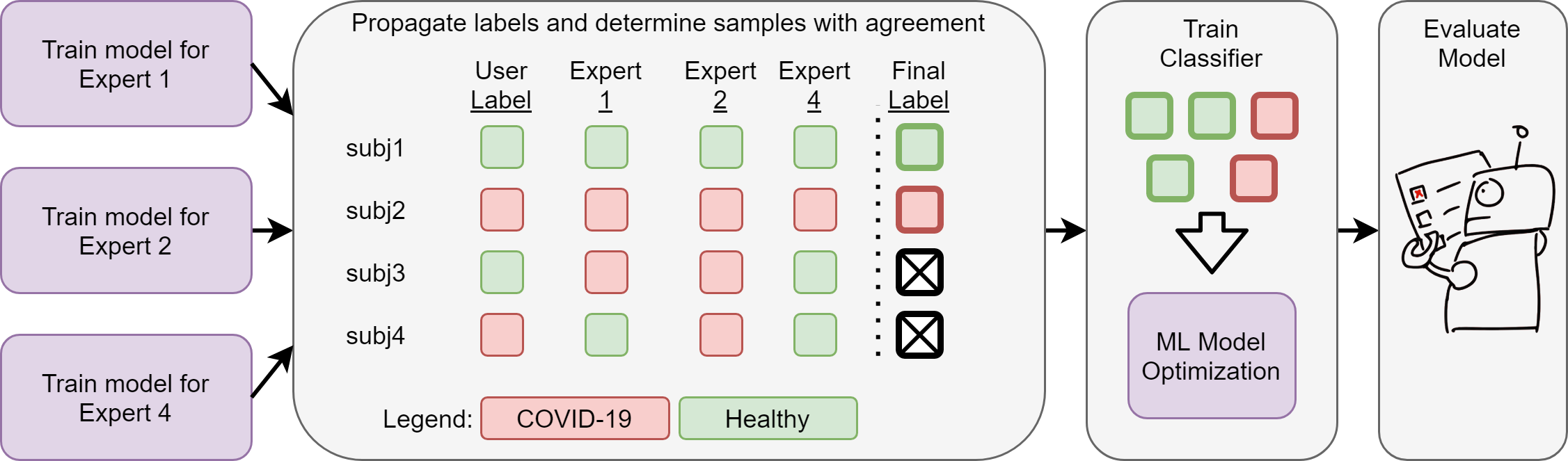}
	\caption{The SSL approach consists of 1) training three separate ML models based on each expert's labels, 2) using the models to classify the unlabeled samples, 3) training a new classifier with the samples exhibiting a significant agreement between the user labels and expert models, and 4) testing the final model.}
	\label{Fig:ssl_pipeline}
\end{figure*}

The semi-supervised learning methodology is illustrated in Fig. \ref{Fig:ssl_pipeline}. First, three distinct ML models were trained based on each expert's COVID-19 versus healthy cough labels and optimized based on the procedure in Section \ref{sec:ml_opt}. Each expert model aggregated the COVID-19 detection probabilities for the coughs of a given recording using the mean logit score in Equation \ref{eq:logit_mean}. Then, the optimal classification threshold of each model was applied on these scores to produce a binary COVID-19 or healthy label for each recording. This procedure resulted in three to four labels per recording: three labels from the expert models, and one user label for the recordings for which this information was provided. In the event that an expert labeled a given recording, the expert's original COVID-19 or healthy diagnosis was maintained rather than the output of the model corresponding to that expert.

Next, a subset of high-confidence samples was identified by comparing the agreement of the expert models and user labels. The recordings with a high degree of agreement for being either COVID-19 positive or healthy were used to train one final classifier, and the rest were discarded. In order to select the final dataset, three different agreement schemes were tested and assessed in terms of database size and class separation:
\begin{enumerate}
    \item \textit{Universal Agreement:} All three expert models have the same label as the user label. This scheme limits the analysis to only the user-labeled datapoints.
    \item \textit{Expert Agreement:} All three expert models have the same label. This scheme bypasses the user label and can thus be applied on unlabeled samples in the dataset.
    \item \textit{Majority Agreement:} Either all three expert models have the same label, or two expert models have the same label as a user. This is the least conservative scheme as it allows one disagreement or missing user label.
\end{enumerate}

\subsection{Model Evaluation}
\label{sec:testing}

Once all of the models described in Section \ref{sec:ml_opt} were trained and optimized, they were tested on the COUGHVID private test set to determine their generalization capabilities on unseen recordings \cite{orlandic_coughvid_2021}. This is a set of 625 recordings that have been labeled by at least one expert, and most recordings contain a user status label. The utilized success metric is AUC because it is a fair metric in terms of class imbalance. For each classification task, the training data labeling scheme was also used for testing. For example, when a model was trained using the annotations of Expert 2, it was tested only on the subset of testing data that had been labeled by Expert 2. In the case of the semi-supervised learning approach, the expert model label propagation procedure described in Section \ref{sec:ssl} was also performed on the testing set, and the final testing samples were identified using the same agreement scheme. Although the labels of the testing data may change between classification tasks, all data is drawn from the same set of recordings.

Once the various success metrics of the classifiers were computed, we assessed the most important features contributing to each classifier's outcome using the Shapley additive explanation (SHAP) values. These  are measures of the relative importance of each feature, indicating which feature domains and specific measures had the greatest influence on the model's decision \cite{Erikstrumbelj2010AnTheory}.

\section{Results}

\subsection{Semi-Supervised Learning Agreement Scheme Selection}

First, we evaluate which agreement scheme among the expert models and user labels, described in Section \ref{sec:ssl}, strikes the optimal trade-off between training dataset coverage and label consistency. The three schemes were applied on the entire database and the number of remaining samples, both recordings and segmented coughs, of each class is reported in Table \ref{tab:agreement_ssl}. This analysis provides an idea of how much training and testing data is maintained, as insufficient data may result in significant overfitting in the final model. Furthermore, the features described in Section \ref{section:feature_extraction} were computed for all of the segmented cough signals in each scheme to assess how the agreement scheme affects the class separation, which is used as a proxy measure of the label consistency across expert models. To quantify this class separation, the Jensen-Shannon divergence of each feature distribution in the COVID-19 and healthy cough classes was computed and averaged across all of the computed features of the training data. This metric ranges from 0 to 1, with higher values corresponding to a larger class separation. These results are displayed in Table \ref{tab:agreement_ssl}, along with the same metrics computed on the user-labeled data subset.

\begin{table}
\centering
\caption{Agreement Scheme Dataset Coverage}
\begin{tabular}{|l|l|l|l|l|l|l|} 
\hline
\begin{tabular}[c]{@{}l@{}}Label\\Scheme\end{tabular}    & \begin{tabular}[c]{@{}l@{}}Training\\Recs.\\(+)\end{tabular} & \begin{tabular}[c]{@{}l@{}}Training\\Coughs\\(+)\end{tabular} & \begin{tabular}[c]{@{}l@{}}Testing\\Recs.\\(+)\end{tabular} & \begin{tabular}[c]{@{}l@{}}Testing\\Coughs\\(+)\end{tabular} & \begin{tabular}[c]{@{}l@{}}Jensen-Shannon\\Divergence\end{tabular}  \\ 
\hline
User                                                     & \begin{tabular}[c]{@{}l@{}}10,850\\(720)\end{tabular}        & \begin{tabular}[c]{@{}l@{}}25,227\\(1,716)\end{tabular}       & \begin{tabular}[c]{@{}l@{}}287\\(163)\end{tabular}          & \begin{tabular}[c]{@{}l@{}}1,098\\(637)\end{tabular}                                                    & 0.00877                                            \\ 
\hline
\begin{tabular}[c]{@{}l@{}}SSL \\Universal\end{tabular} & \begin{tabular}[c]{@{}l@{}}2325\\(14)\end{tabular}           & \begin{tabular}[c]{@{}l@{}}5515\\(45)\end{tabular}            & \begin{tabular}[c]{@{}l@{}}28\\(2)\end{tabular}             & \begin{tabular}[c]{@{}l@{}}104\\(10)\end{tabular}                                                       & 0.0954                                             \\ 
\hline
\begin{tabular}[c]{@{}l@{}}SSL \\Expert\end{tabular}    & \begin{tabular}[c]{@{}l@{}}4128\\(98)\end{tabular}           & \begin{tabular}[c]{@{}l@{}}9583\\(295)\end{tabular}           & \begin{tabular}[c]{@{}l@{}}141\\(12)\end{tabular}           & \begin{tabular}[c]{@{}l@{}}501\\(62)\end{tabular}                                                       & 0.0519                                             \\ 
\hline
\begin{tabular}[c]{@{}l@{}}SSL \\Majority\end{tabular}  & \begin{tabular}[c]{@{}l@{}}8331\\(285)\end{tabular}          & \begin{tabular}[c]{@{}l@{}}20337\\(848)\end{tabular}          & \begin{tabular}[c]{@{}l@{}}240\\(53)\end{tabular}           & \begin{tabular}[c]{@{}l@{}}876\\(239)\end{tabular}                                                        & 0.0284                                             \\
\hline
\end{tabular}
\label{tab:agreement_ssl}
\end{table}

As Table \ref{tab:agreement_ssl} shows, in the universal agreement scheme, there were only 14 COVID-19 positive recordings remaining in the training dataset.  This amount is insufficient for the model to perform generalization. As expected, the majority agreement scheme produces the largest number of training samples. Intuitively, the Jensen-Shannon divergence decreases as the agreement scheme gets less conservative, meaning that the universal agreement scheme exhibits the largest class separation across features while the majority scheme has less pronounced differences between features. However, this increase in class separation comes at the expense of a decrease in dataset coverage, so the method that conserves the most data is maintained. This decision is in line with the findings of Zhu et al., which noted that SSL schemes prioritizing a larger initial dataset with moderately consistent labels performed better than small datasets with very reliable labels \cite{zhu_speech_2021}. 

The majority agreement scheme was selected to identify the final COVID-19 and healthy cough samples. While this scheme has the smallest class separation of the other semi-supervised learning schemes, its Jensen-Shannon divergence is still more than three times higher than that of the user-labeled scheme. Furthermore, the number of training samples used in this agreement scheme is only 23\% smaller than that of the user-labeled scheme, meaning that the increase in class separability does not sacrifice much of the data coverage. The percentage of COVID-19-labeled coughs in the majority agreement scheme is 3.3\%, which is lower than the 6.2\% in the user-labeled dataset. However, this class imbalance is handled in training by applying the SMOTE method described in Section \ref{sec:ml_opt}.

\subsection{Intra-Class Consistency Analysis}

\begin{table}
\centering
\caption{Final Model Selection}
\begin{tabular}{|l|l|l|} 
\hline
Label Type & Model Used                   & Hyperparameters                                                                                                                        \\ 
\hline
User       & Linear Discriminant Analysis & None                                                                                                                                   \\ 
\hline
Exp. 1     & Logistic Regression          & \begin{tabular}[c]{@{}l@{}}C=18.31,\\class\_weight=None,\\solver=`newton-cg'\end{tabular}          \\ 
\hline
Exp. 2     & Logistic Regression          & \begin{tabular}[c]{@{}l@{}}C=0.01038,\\class\_weight=`balanced’,\\solver=`newton-cg'\end{tabular}  \\ 
\hline
Exp. 4     & Logistic Regression          & \begin{tabular}[c]{@{}l@{}}C=0.3306,\\class\_weight=None,\\solver='lbfgs'\end{tabular}             \\ 
\hline
SSL        & Logistic Regression          & \begin{tabular}[c]{@{}l@{}}C=0.01009,\\class\_weight=`balanced’,\\solver=`newton-cg'\end{tabular}  \\
\hline
\end{tabular}
\label{tab:model_opt_results}
\end{table}

Once all three expert models were trained and optimized using the procedure in Section \ref{sec:ml_opt}, these labels were propagated onto both the training and testing datasets. By selecting the subset of recordings for which the majority of labels were in agreement, we expanded the expert knowledge, combined with user self-report labels, to identify training and testing samples that had a high probability of having correct labels. The final optimized models evaluated in this work are displayed in Table \ref{tab:model_opt_results}, complete with their respective hyperparameters selected through TPE.

\begin{figure}
  \centering
  \begin{tabular}{@{}c@{}}
    \includegraphics[width=\linewidth]{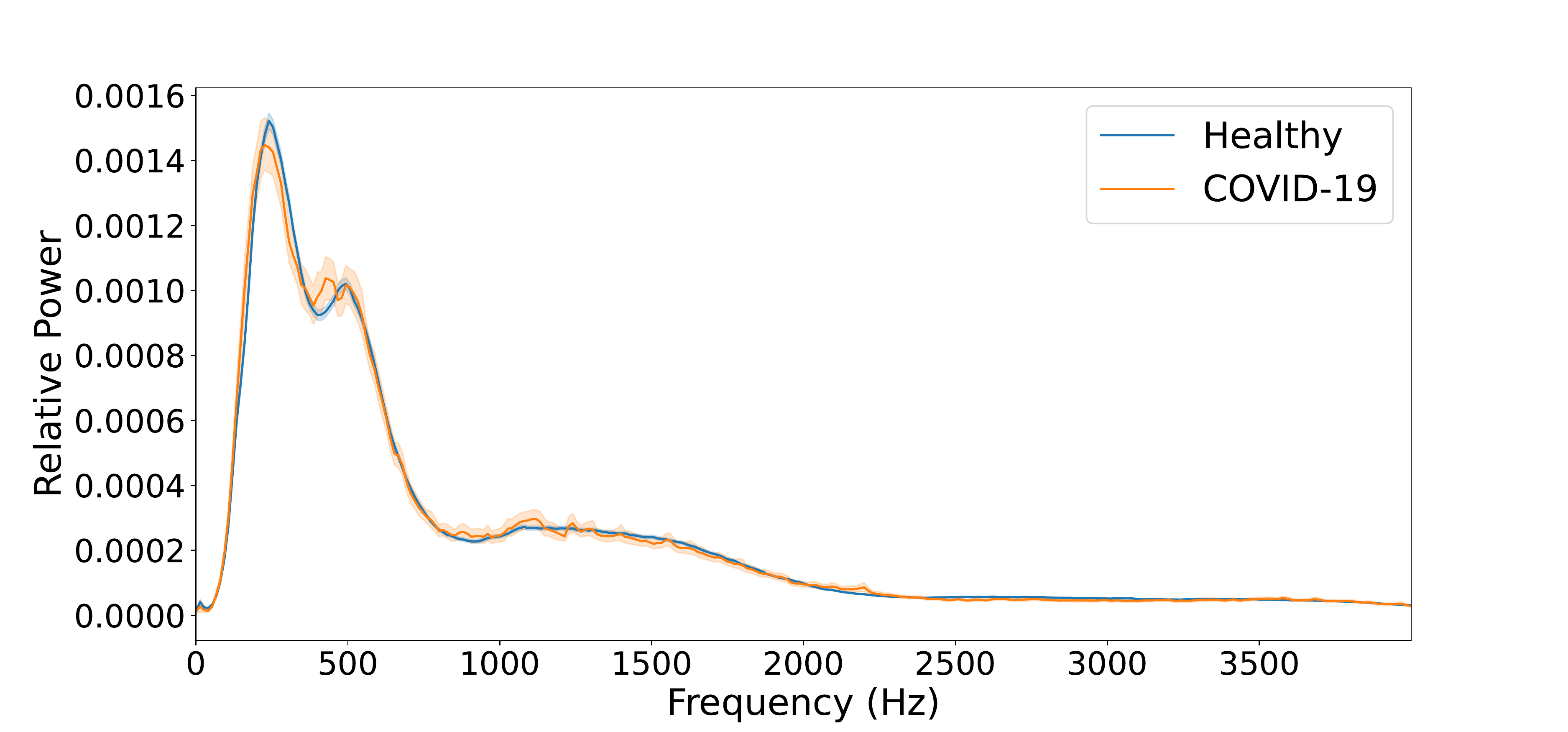} \\[\abovecaptionskip]
    \small (a) User labeling scheme
  \end{tabular}

  \vspace{\floatsep}

  \begin{tabular}{@{}c@{}}
    \includegraphics[width=\linewidth]{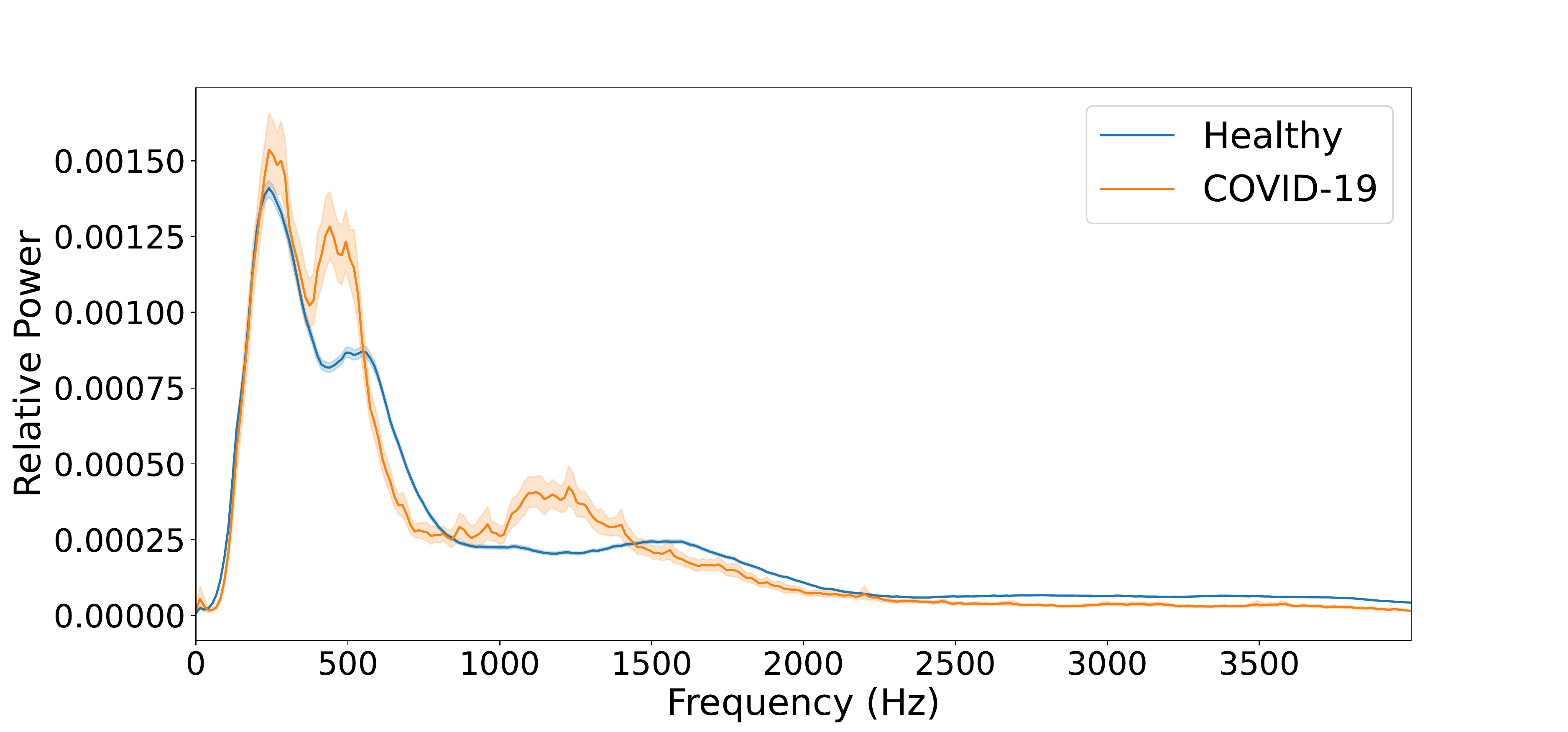} \\[\abovecaptionskip]
    \small (b) SSL labeling scheme
  \end{tabular}

  \caption{Average normalized PSD of all cough signals in the training dataset belonging to each class according to the user labels and SSL labels.}\label{fig:psd}
\end{figure}

To assess the difference in audio properties of coughs labeled as COVID-19 and healthy in this new dataset compared to those of the user labels, the average normalized PSD curves of cough signals belonging to each class are plotted in Figures \ref{fig:psd}a and \ref{fig:psd}b.
The figures show a solid line, indicating the average PSD, as well as the 95\% confidence interval across all segmented cough audio samples of a given class in each labeling scheme. Fig. \ref{fig:psd}a shows very few differences between the spectra of the user-labeled COVID-19 and healthy coughs, with a slight variation in the 400-550 Hz and 1000-1500 Hz ranges. In comparison, Fig. \ref{fig:psd}b depicts a similarly shaped spectrum of healthy coughs as the user-labeled healthy coughs, but there are much more pronounced differences between the COVID-19 coughs and healthy ones. The bandpowers of the COVID-19 coughs are significantly higher in the 400-550 Hz and 1000-1500 Hz ranges than those of healthy coughs, with p-values of $1.4 \times 10^{-36}$ and $1.2 \times 10^{-64}$, respectively. This analysis highlights the substantial difference in spectral features of COVID-19 and healthy coughs identified through the semi-supervised learning approach. It is also consistent with the findings of Table \ref{tab:agreement_ssl}, which shows that on average, the chosen SSL dataset exhibits over 3x more class separability than the user-labeled data in terms of the average Jensen-Shannon divergence across all extracted audio features.

To expand on the feature analysis, the top five most important features for each classifier, determined by the SHAP values of each classifier, are displayed in Table \ref{tab:shap}. 
\begin{table}
\centering
\caption{SHAP Feature Ranking across Classifiers}
\begin{tabular}{|l|l|l|l|l|l|} 
\hline
\begin{tabular}[c]{@{}l@{}}Feature\\Ranking\\(SHAP)\end{tabular} & User                                                        & Expert 1                                                   & Expert 2                                                  & Expert 4                                                  & SSL                                                    \\ 
\hline
1                                                                & \begin{tabular}[c]{@{}l@{}}EEPD\\350-400\end{tabular}    & \begin{tabular}[c]{@{}l@{}}Spectral\\Centroid\end{tabular} & Gender                                                    & \begin{tabular}[c]{@{}l@{}}Crest\\Factor\end{tabular}     & \begin{tabular}[c]{@{}l@{}}MFCC~\\Std. 0\end{tabular}  \\ 
\hline
2                                                                & \begin{tabular}[c]{@{}l@{}}EEPD\\600-650\end{tabular}    & \begin{tabular}[c]{@{}l@{}}RMS\\Power\end{tabular}         & \begin{tabular}[c]{@{}l@{}}MFCC\\Mean 7\end{tabular}      & \begin{tabular}[c]{@{}l@{}}MFCC\\Mean 0\end{tabular}      & \begin{tabular}[c]{@{}l@{}}MFCC\\Mean 1\end{tabular}   \\ 
\hline
3                                                                & \begin{tabular}[c]{@{}l@{}}RMS\\Power\end{tabular}          & \begin{tabular}[c]{@{}l@{}}MFCC\\Std. 0\end{tabular}       & \begin{tabular}[c]{@{}l@{}}PSD\\550-800\end{tabular}   & \begin{tabular}[c]{@{}l@{}}Spectral\\Slope\end{tabular}   & \begin{tabular}[c]{@{}l@{}}MFCC\\Mean 9\end{tabular}   \\ 
\hline
4                                                                & \begin{tabular}[c]{@{}l@{}}EEPD\\900-950\end{tabular}    & \begin{tabular}[c]{@{}l@{}}Spectral\\Spread\end{tabular}   & \begin{tabular}[c]{@{}l@{}}MFCC\\Std. 5\end{tabular}      & \begin{tabular}[c]{@{}l@{}}Spectral\\Rolloff\end{tabular} & Gender                                                 \\ 
\hline
5                                                                & \begin{tabular}[c]{@{}l@{}}Dominant\\Frequency\end{tabular} & \begin{tabular}[c]{@{}l@{}}Spectral\\Skewness\end{tabular} & \begin{tabular}[c]{@{}l@{}}EEPD~\\400-450\end{tabular} & \begin{tabular}[c]{@{}l@{}}MFCC\\Mean 9\end{tabular}      & \begin{tabular}[c]{@{}l@{}}MFCC\\Mean 7\end{tabular}   \\
\hline
\end{tabular}
\label{tab:shap}
\end{table}
When we analyze the three expert models, it is clear that there are few features in common between the classifiers and they each weigh features of different domains (i.e., time, frequency, and Mel frequency) with varying importance. The semi-supervised learning classifier, on the other hand, has features in common with several expert models (MFCC standard deviation 0, MFCC mean 7, and gender). Furthermore, the majority of its top features are in the Mel frequency domain, which is meant to model how the human auditory system processes sound signals. 

\subsection{Open-Sourced SSL Dataset}

In order to contribute to further research in the field of COVID-19 cough sound diagnosis, we have added the training labels obtained through our SSL majority agreement scheme to the latest version of the  \href{https://zenodo.org/record/7024894#.YwjMAXZByUk}{COUGHVID dataset public Zenodo repository}. This version has been expanded to include all of the crowdsourced recordings obtained through October 2021, whereas the original dataset only contained recordings uploaded through December 2020. These new labels can be found in the newly added \texttt{status\_SSL} column of the \texttt{metadata\_compiled} CSV file.

The new SSL scheme provides labels for training 1,018 recordings that were previously unlabeled by users or experts, which demonstrates the utility of SSL in utilizing data that had previously been unusable. Furthermore, there are 581 recordings that the users labeled with the ambiguous ``symptomatic" label, but the SSL model provides a ``COVID-19" or ``healthy" label. A mere 32 of these coughs were labeled by the SSL model as COVID-19 positive, which is feasible considering the COVID-19 infection rates during the period of recording. 

Users of the COUGHVID dataset can use these new labels and corresponding data samples to augment their COVID-19 cough classification models with highly consistent training data. The same SSL label expansion procedure was conducted for the private testing dataset described in \ref{sec:testing}, so users are welcome to test their models against these labels as ground-truth, but must acknowledge that these labels are not confirmed by RT-PCR tests.

\subsection{ML Model Evaluation}

\begin{table}
\centering
\caption{Model Testing Results}
\begin{tabular}{|l|l|l|l|l|l|l|} 
\hline
Model           & CV AUC & \begin{tabular}[c]{@{}l@{}}Test AUC\\(Not Agg.)\end{tabular} & \begin{tabular}[c]{@{}l@{}}Test AUC\\(Agg.)\end{tabular}    \\ 
\hline
User   & 0.591  & 0.564                                                              & 0.562                                                         \\ 
\hline
Exp. 1        & 0.653  & 0.652                                                              & 0.681                                                          \\ 
\hline
Exp. 2        & 0.669  & 0.663                                                              & 0.743                                                          \\ 
\hline
Exp. 4        & 0.644  & 0.561                                                              & 0.593                                                         \\ 
\hline
SSL & 0.883  & 0.763                                                              & 0.797                                                         \\
\hline
\end{tabular}
\label{tab:final_results}
\end{table}

To demonstrate how the SSL dataset can be used to train cough classification ML models, one final model was developed using the procedure in \ref{sec:ml_opt} using the SSL labels. The final testing results of each model are displayed in Table \ref{tab:final_results}, which shows all of the accuracy metrics, as well as the AUC obtained during cross-validation, non-aggregated testing (i.e., testing on every individual cough sound), and aggregated testing on each recording using one of the formulas in Equations \ref{eq:logit_mean} and \ref{eq:logit_median}. We observe an average 5.26\% increase in accuracy between non-aggregated and aggregated testing. This implies that testing each cough separately and combining the results for each recording enhances the model's performance. Aggregating the probabilities of each cough sound in a recording may exploit the correlations between the coughs and diminish the effects of outlier cough sounds, thus providing a more robust classification than predicting each cough sound separately.

The model trained on self-reported user data exhibited the worst performance. Reaching only  a testing AUC of 0.562, it was scarcely better than a random classifier. We observe a high variance in the success of the expert models, with Expert 2 having the highest AUC and Expert 4 the lowest. The AUC of the semi-supervised classification method is, on average, 15.6\% higher than that of the expert models, and 29.5\% higher than that of the user model. 

\begin{figure}[ht]
  \centering
  \includegraphics[width=0.8\linewidth]{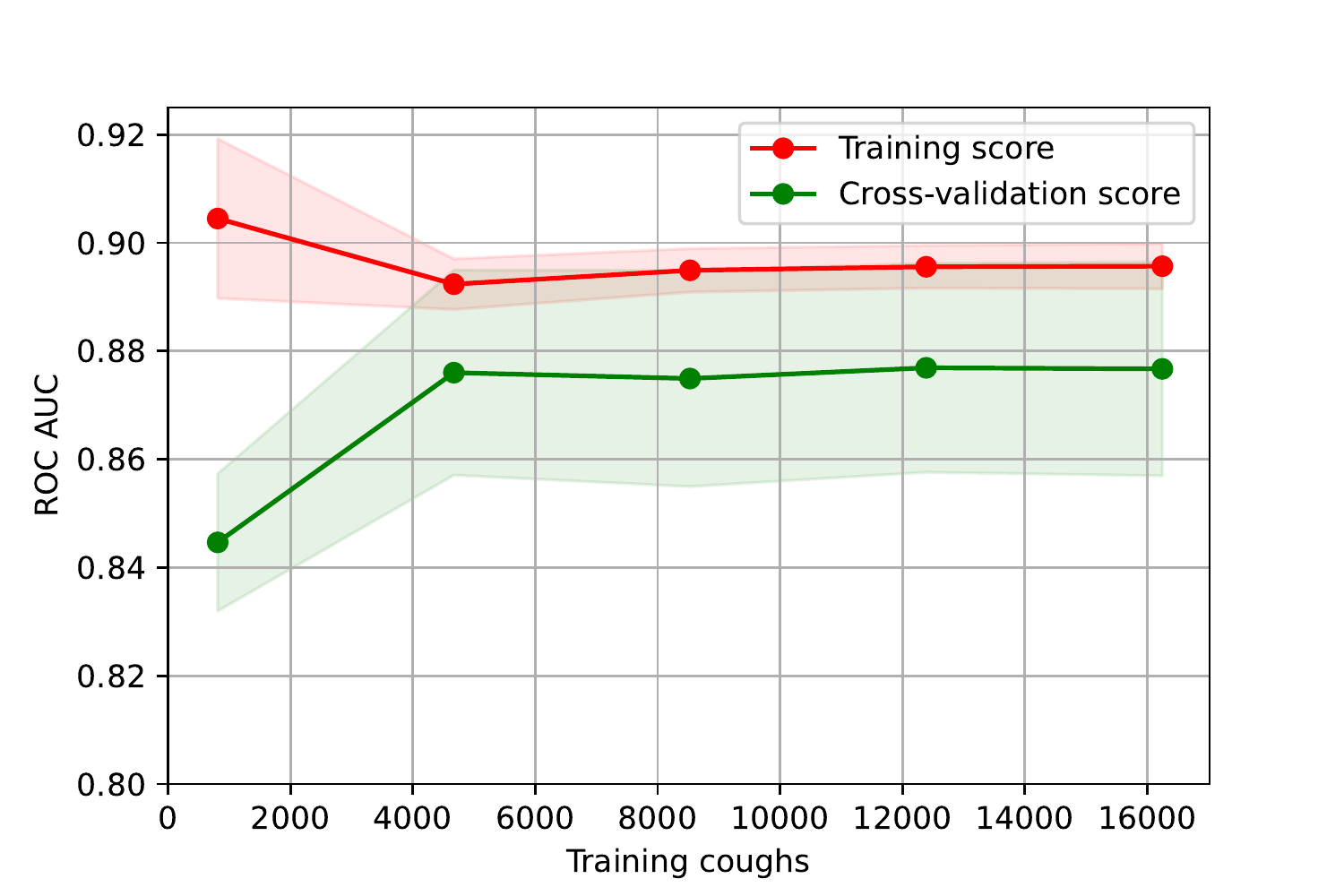} 
\caption{Learning curve of the final optimized SSL classifier displaying the training and cross-validation accuracy using varying sizes of the training dataset.}
\label{fig:learning_curve}
\end{figure}

The final SSL model utilized 32 of the available 66 features, which were selected through RFECV. The learning curve of the final SSL model is displayed in Fig. \ref{fig:learning_curve}, which shows the effect of varying the training data size (in terms of number of segmented coughs) on the training and validation accuracy. The solid line depicts the mean scores across the five CV folds, and the shading around the line indicates one standard deviation from the mean. We can see that the model converges to a  validation accuracy at around 4,000 training coughs, indicating that the model has sufficient training samples to gain insights from the features. Furthermore, the relatively small 2\% gap between the training and validation scores indicates that the variance of the model is low, meaning that it is not over-fitting on the training samples.

\begin{figure}[ht]
  \centering
  \includegraphics[width=0.8\linewidth]{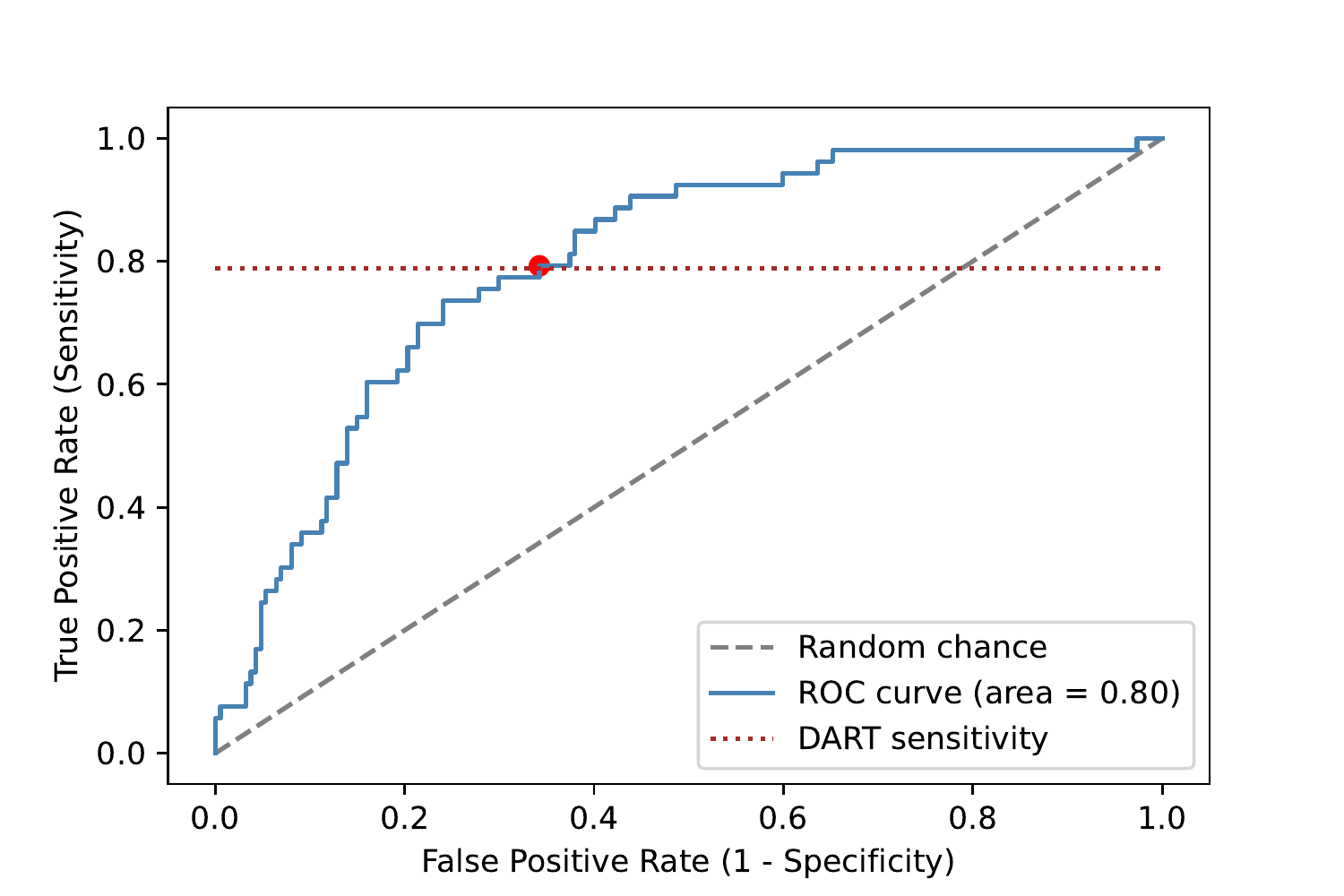} 
\caption{The final ROC curve of the semi-supervised classifier on the aggregated testing set. At the red point, the model achieves a sensitivity of 79.2\%, which is comparable to that of DART COVID-19 tests, and a specificity of 65.8\%.}
\label{fig:roc}
\end{figure}

The ROC curve of the semi-supervised learning classifier is displayed in Fig. \ref{fig:roc}. The model had an AUC of 0.797 on the private testing set. To evaluate the classification accuracy of such a model, we compare its sensitivity and specificity to those of commonly-used at-home COVID-19 tests. The Direct Antigen Rapid Test (DART) for COVID-19 screening was reported to have a sensitivity of 78.9\% and specificity of 97.1\% within 0 to 12 days from symptom onset \cite{harmon_validation_2021}. To achieve a comparable sensitivity, our classifier exhibits a specificity of 65.8\%, which is a 32.8\% decrease from that of DART tests. This decrease in specificity could be justified by the inexpensive, ubiquitous, and non-invasive nature of an audio-based screening tool versus the traditional nasal swab.

\section{Discussion}

By integrating the knowledge from three medical experts with the user labels through the Pseudo-Label-based semi-supervised learning approach, we identified a subset of cough recordings that had a high probability of belonging to each class. We see from the averaged PSD curves of COVID-19 and healthy coughs in Fig. \ref{fig:psd}b that the majority-voting approach between the expert pseudo-labels successfully identified two classes of coughs with significantly different spectral characteristics and a high class separation in the extracted audio features. As each expert model and user labels aimed to separate COVID-19 versus healthy coughs, it can be postulated that these spectral characteristics are present in the underlying distributions of the two classes of cough sounds. Furthermore, these spectral differences were much less pronounced when the same analysis was performed for the user-labeled data in Fig. \ref{fig:psd}a. These figures illustrate the fact that the class separation was over three times higher in the SSL training data than of the user labeling in terms of the Jensen-Shannon divergence of the feature distributions. This increase in class separation did not come at a significant cost to the data coverage, as the training data size only decreased by 23\%.  

An analysis of the most important features of the expert and semi-supervised classifiers in Table \ref{tab:shap} revealed that there were few important features common among all expert classifiers, which may be reflective of the lack of expert agreement observed in \cite{orlandic_coughvid_2021}. This table also showed that the semi-supervised approach learned which features were important from most of the expert models and weighed these features heavily in its own classification outcome. Furthermore, the fact that the semi-supervised model relies almost entirely on MFCC features might imply that the classifier is learning to model the human auditory system, since these features model how humans perceive sound.

Finally, an analysis of the model testing results in Fig. \ref{tab:final_results} reveals the drawbacks of classic supervised learning approaches, as well as the improved performance of SSL. First, we note that the model that was trained and tested on crowdsourced user labels achieved a testing AUC of 0.562, and such a low score implies that there is significant mislabeling present in the dataset. Next, we note a wide variance between the success of each expert model, with the AUC scores ranging from 0.593 to 0.743. This means that the expert labels tend to be inconsistent between and even within each expert's labels. However, despite the setback of label ambiguity, the semi-supervised modeling approach achieved a high final AUC score of 0.797, which was at least 7.3\% higher than any of the expert models. These results indicate that integrating the medical knowledge of multiple experts in a semi-supervised fashion results in a more robust, consistent classifier than supervised learning based on any of the individual experts' labels.

Although the proposed SSL method showed an increased model performance on the COUGHVID hidden test set, such an approach must be thoroughly validated on PCR-confirmed cough samples. Furthermore, the algorithm is unable to account for concept drift due to the varying symptomologies of the different COVID-19 virus variants. The data used in training was obtained through October 2021, whereas the Omicron variant -- which had a significantly lower rate of respiratory symptoms than previous variants -- was first reported in November 2021 \cite{bouzid_comparison_2022}. Therefore, these issues must be addressed in future work to accurately assess the clinical usefulness of such respiratory disorder classification models.

\section{Conclusion}

Labeling medical data requires significant time and effort from expert annotators. Moreover, this tedious process often leads to inconsistencies due to a lack of agreement between experts. This situation is a key drawback to confront new viruses, as it has happened with COVID-19. In this work, we have shown that using an SSL model development method, it is possible to overcome expert label scarcity and inconsistency -- as well as user mislabeling of crowdsourced medical datasets -- to identify a subset of data points with a high-class separability. We applied this approach for the first time to the task of COVID-19 screening from cough audio recordings and achieved a performance increase of 15.6\% from fully-supervised expert models and 29.5\% from crowdsourced user labeling.  We have also achieved a 3x increase in class separability in the training data identified through semi-supervised learning compared to user labeling, while only decreasing the database coverage by 23\%. Furthermore, this high-class separation can be associated with specific spectral components of the audio signals, which makes the models explainable from an acoustic perspective.

Using the COUGHVID crowdsourcing dataset, we developed a signal processing pipeline that extracts state-of-the-art audio features from each segmented cough signal. Then, a model optimization procedure fostering generalizability was employed to develop classifiers based on each expert's labels as well as user labels. The expert classifiers were tested on the whole dataset, and recordings whose majority opinion between the expert models and user label agreed were maintained in training the final classifier, which enabled the usage of unlabeled samples. Such an approach can be employed for other respiratory disorder diagnosis tasks based on partially-labeled audio data, thus minimizing the need for excessive labeling from one expert and rather having several experts label small portions of the dataset. This way, the experts' opinions can be combined to account for labeling inconsistencies.

Our proposed semi-supervised learning approach has proven to have the highest classifier performance of all of the tested models, with a final testing AUC of 0.797, which leads to a sensitivity of 79.2\% and specificity of 65.8\%. Using recursive feature elimination, we selected a subsample of 32 features out of the original 66 for training the final model, thereby increasing the model robustness by removing redundant or unnecessary features. SHAP analysis revealed that three of the five most important features of the final semi-supervised classifier were also significant in each expert classifier, which implies that the model successfully integrated each expert's medical knowledge. Such an approach is not specific to the COUGHVID dataset and can be used in any medical database that is only partially labeled by a set of experts. 

While the class separation is significant between the COVID-19 and healthy cough recordings identified through the expert pseudo-label comparison, we cannot be sure that these trends truly apply to COVID-19 positive and healthy coughs unless they are tested on cough recordings from RT-PCR-confirmed COVID-19 positive and negative individuals. However, in the absence of an extensive, RT-PCR-validated dataset, our proposed approach can be easily used to improve the quality of large, crowdsourced COVID-19 cough databases and identify samples with consistent patterns in the cough recordings of each class. These re-labeled coughs can then be used to augment datasets of medically confirmed cough sounds to enhance the training data size and potentially improve the classification accuracy.

\section{Acknowledgements}
This project has received funding from the European Union’s Horizon 2020 research and innovation programme under grant agreement No 101017915 (DIGI-PREDICT), as well as the DeepHealth Project (GA No. 825111) and the Swiss NSF ML-Edge Project (GA No. 182009). T. Teijeiro is supported by a Maria Zambrano fellowship (MAZAM21/29) from the University of Basque Country and the Spanish Ministry of Universities, funded by the European Union-Next-GenerationEU.

\bibliographystyle{IEEEtran}
\bibliography{refs}


\end{document}